\documentclass[twocolumn,showpacs,preprintnumbers,draftclsnofoot]{revtex4}
\usepackage{amsmath}
\usepackage{amssymb}
\usepackage{graphicx}
\usepackage{dcolumn}
\usepackage{bm}
\usepackage{amssymb}
\usepackage{graphicx}
\usepackage{dcolumn}
\usepackage{bm}
\begin{document}

\title{ Goos-H$\ddot{a}$nchen shifts of bilayer meta-grating with unidirectional guide resonance }
\author{Zhihao Xu and Ma Luo\footnote{Corresponding author:swym231@163.com} }
\affiliation{School of Physics and Optoelectronic Engineering, Guangdong Polytechnic Normal University, Guangzhou 510665, China}

\begin{abstract}

Bilayer meta-gratings with asymmetric structural parameters could host unidirectional guide resonances. The distribution of unidirectional guide resonances in the space of structural parameters and synthetic parameters is identified. As the incident optical beam being resonant with the unidirectional guide resonance, the Goos-H$\ddot{a}$nchen shifts of the scattered beams exhibit two anomalous behaviors: the resonant peak of the Goos-H$\ddot{a}$nchen shift is accompanied by constant transmittance and reflectance; the magnitude of the Goos-H$\ddot{a}$nchen shift is not always proportional to the quality factor of the unidirectional guide resonance. The temporal coupled mode theory analysis reveals that the first anomalous behavior is due to interference between direct scattering and radiation from the unidirectional guide resonance; the Goos-H$\ddot{a}$nchen shifts are proportional to the group velocity as well as the quality factor of the unidirectional guide resonance. Numerical simulations of incidence of Gaussian beam with finite beam width provide intuitive visualization of the Goos-H$\ddot{a}$nchen shift.

\end{abstract}

\maketitle

\section{Introduction}

Meta-gratings with subwavelength periodic structure offer numerous means to manipulate light wave \cite{metagrating}. When the frequency and lateral wavenumber of an incident light wave are scanned across the band structure of the leaky resonant modes of the meta-grating, the scattered waves exhibit resonant peaks in the spectrum of reflectance and transmittance, as well as lateral beam shifting designated as Goos-H$\ddot{a}$nchen (GH) shift \cite{firstGH47}. By utilizing the GH shift to manipulate light, varying type of applications has been proposed, such as sensors \cite{GHapplSensing1,GHapplSensing2}, optical information storage \cite{GHapplStore1,GHapplStore2,GHapplStore3}, wavelength division de-multiplexers \cite{GHapplMulti}, optical switches \cite{GHapplSwitch}, and polarization beam splitters \cite{GHapplBeam}. Thus, various resonant microstructures have been proposed to enhance the GH shift, including interface with Brewster effects \cite{Brewster1,Brewster2,Brewster3}, surface plasmon polaritons \cite{sppGH1,sppGH2,sppGH3,sppGH4,sppGH5}, Fabry-Perot resonances \cite{fpcavityGH1,fpcavityGH2,fpcavityGH3,fpcavityGH4}, Bloch surface waves \cite{blochGH1,blochGH2}, Tamm plasmon polaritons \cite{tamnGH1,tamnGH2,tamnGH3}, and quasi-bound states in the continuum (quasi-BICs) \cite{bicGH1,bicGH2,bicGH3}. Among the numerous schemes to enhance GH shift, the superiority of all dielectric meta-grating is the absence of absorption loss. As a result, the GH shift of all dielectric meta-grating have attracted tremendous attention in recent years.

Varying types of topological photonic modes could lay among the band structure of the leaky resonant modes, such as bound states in the continuum (BICs) and unidirectional guide resonance (UGR) \cite{raDongJianWen}. BICs is a type of singular mode without radiative loss, so that the quality factor (Q factor) is infinitely large \cite{bicItSelf1,bicItSelf2,bicItSelf3,bicItSelf4}. Within the band structure of the leaky resonant modes, the distribution of the far-field radiative polarization forms a vortex around the BIC, so that the magnitude of the far-field radiation of the BIC is zero \cite{bicVortex}. The leaky resonant modes near to the BICs are quasi-BICs with finite but ultra-high Q factor \cite{quasiBIC01,quasiBIC02,quasiBIC03,quasiBIC04,quasiBIC05,quasiBIC06,quasiBIC07,quasiBIC08,quasiBIC09,quasiBIC10,quasiBIC11,quasiBIC12,quasiBIC13,quasiBIC14,quasiBIC15,quasiBIC16,quasiBIC17,quasiBIC18,quasiBIC19,quasiBIC20,quasiBIC21}. The quasi-BICs induce Fano-resonant with sharp peak in the spectrum of reflectance or transmittance, which is accompanied with large enhancement of GH shift \cite{bicGH1}. According to the stationary-phase theory \cite{stationaryPt48}, the GH shift is determined by the slope of the reflection (transmission) phase. For the quasi-BICs with large Q factor, the slope of the reflection (transmission) phase is proportional to the Q factor, so that the maximum magnitude of the GH shift is proportional to the Q factor \cite{bicGH2}.

When the meta-gratings have bilayer structure, the lateral offset between the two layers breaks the up-down mirror symmetry of the structure. Thus, the far-field radiative polarizations of the up and down backgrounds become different. In the other words, the asymmetric structures induce radiation asymmetry, whose radiative loss to the up and down background have different magnitude or phase. The vortices of the far-field radiative polarizations of the up and down backgrounds could locate at different points in the parameter space. In the center of the vortices, the far-field radiation to one side of the meta-grating is zero, while the far-field radiation to the other side is finite, so that the modes are designated as UGRs \cite{cPointFromV1,cPointFromV3}. When the structures preserve center inversion symmetry, the UGRs appear in pair in the synthetic parameter space, which is consisted of the lateral offset of the bilayer structure and the lateral wavenumber of the leaky resonant modes. The two UGRs have opposite lateral wavenumber. If the structures break center inversion symmetry, the pair of UGRs move toward different direction in the synthetic parameter space. In some regime of the structural parameter space, one UGR of the pair of UGRs vanishes.

In this paper, the optical responses of the incident light waves with frequency being equal to the resonant frequency of a UGR, and lateral wavenumber being scanned across the wavenumber of the UGR, have been numerically and theoretically studied. With the same sign of the lateral wavenumber, incident light waves from up and down backgrounds have the same spectrum of reflectance and transmittance, but different GH shifts. In contrast to the GH shift driven by quasi-BICs, numerical simulations showed that the GH shifts driven by the UGRs exhibit anomalous behaviors. The spectrum of transmittance is near to constant unity, but the spectrum of the transmission phase has large slope, so that the GH shift is accompanied by a constant transmittance. For some of the UGRs, the magnitude of the GH shift is proportional to the Q factor of the UGRs, but for the other UGRs, the magnitude of the GH shift is suppressed to a smaller value. In some cases, the GH shift is suppressed to be zero. The temporal coupled mode theory (TCMT) is applied to analyze the numerical results of the GH shifts driven by UGRs \cite{tcmt01,tcmt02,tcmt03}. The analytical solution of transmission (reflection) coefficient is obtained, which in turn gives the expression of the GH shift. The solution shows that the GH shift is proportional to the group velocity as well as the Q factor of the UGR. In addition, the numerical simulation of incident Gaussian beam with finite beam width exhibits that the GH shift is driven by the excitation of the UGR.

The organization of this paper is as follows. In Sec. II, we systematically analyze the distribution of UGRs in structural and synthetic parameter space of the bilayer meta-grating. In Sec. III, the GH shift of the meta-gratings is studied by stationary-phase theory, analyzed by the TCMT, and numerically visualized by simulation of incident Gaussian beam. Finally, in Sec. IV the conclusion is given.

\section{UGR of bilayer meta-grating}

\subsection{Structure of the bilayer meta-grating}

The structure of one unit cell of the bilayer meta-grating is given in Fig. 1(a), which is consisted of periodical Si grating along x axis in the background of SiO$_{2}$, and is uniform along y axis, and has finite thickness along z axis. The period is $a$. The refractive index of Si and SiO$_{2}$ are assumed to be $n_{\text{Si}}=3.4767$ and $n_{\text{SiO}_2}=1.444\equiv n_b$, respectively. The upper and lower gratings of Si is separated by the distance $G$. The thickness of the upper and lower gratings is both $t$, and the width of the slots in the two layers are $w_1$ and $w_2$, respectively. The lateral offset between the slots in the upper and lower layers is $\Delta g$.
The leaky resonant modes satisfy the Bloch periodic boundary condition along x direction, so that the Bloch wave vector along x axis is also period, i.e.,
$k_x \in [-\pi/a, \pi/a]$.
The lateral offset $\Delta g$ and $k_x$ are both periodic, which form synthetic parameter space; $w_1$ and $w_2$ form structural parameter space. In this paper, the other three structural parameters are fixed to be: $a=1000\ \text{nm}$, $G=45\ \text{nm}$, $t=355\ \text{nm}$; the distribution of UGR in the parameters space formed by $\Delta g$, $k_x$, $w_1$ and $w_2$ is studied.
As $\Delta g=0$ or $\Delta g=a/2$, the structure has mirror symmetry about the vertical axis along z axis. As $w_1=w_2$, the structure has center symmetry. If $w_1=w_2=0$, the structure degrades into a bilayer dielectric slab, which host waveguide modes. As the Bloch periodic boundary being applied, the dispersion relationships of the waveguide modes are folded into the first Brillouin Zone.
As $w_1$ and $w_2$ become nonzero, the periodic slots in the two layers bring perturbation to the waveguide modes, so that the waveguide modes are transferred into leaky resonant modes. For the modes above the light cone, the field pattern at far-field region above and below the meta-grating are radiative plane waves that carry energy loss to the top and bottom background, respectively.
The coefficients of the radiative plane waves in the top and bottom background are $c_{\text{up}}$ and $c_{\text{down}}$, respectively. The total magnitude of the two coefficients can be normalized by the stored energy of the leaky resonant mode in the grating region, while the relative magnitude and phase between $c_{\text{up}}$ and $c_{\text{down}}$ feature the radiation asymmetry of the leaky resonant mode.
The directionality and radiative phase difference of the leaky resonant mode are defined as
$\eta = \frac{|c_{\text{up}}|^2 - |c_{\text{down}}|^2}{|c_{\text{up}}|^2 + |c_{\text{down}}|^2}$
and
$\delta = \text{Arg}\left(\frac{c_{\text{up}}}{c_{\text{down}}}\right)$
respectively.

\begin{figure*}[tbp]
\scalebox{0.53}{\includegraphics{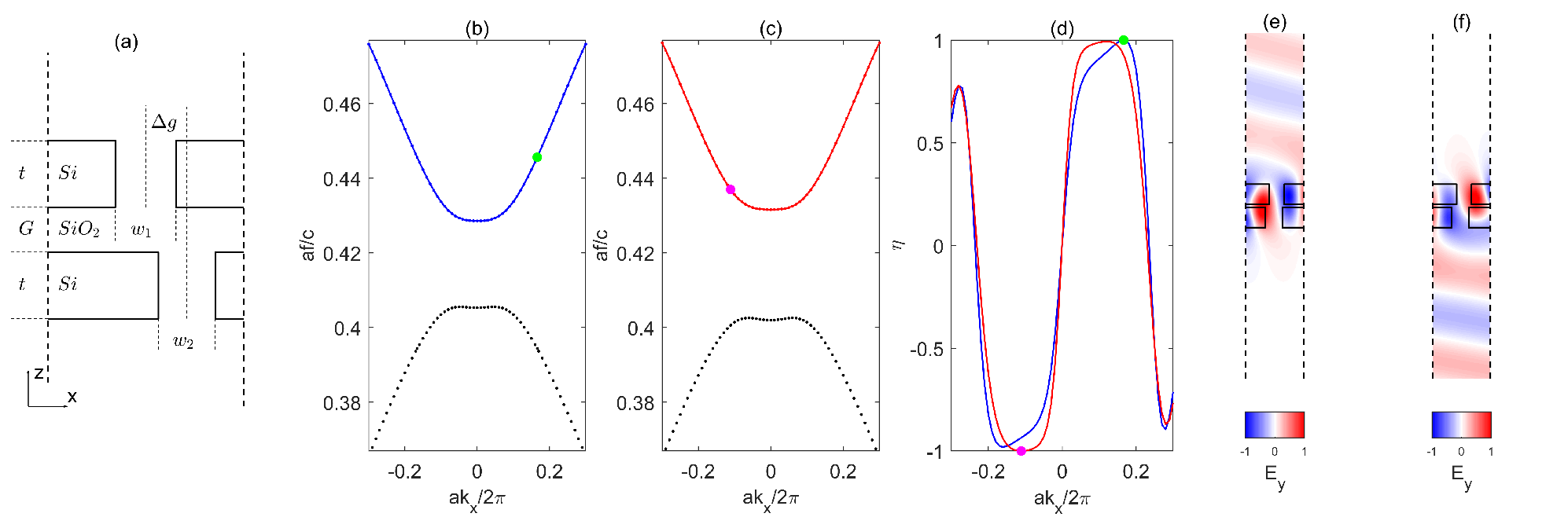}}
\caption{ (a) Schematic of the structure of the bilayer meta-grating in one period. The band structure of the leaky resonant modes as $\Delta g=0.053a$ and $\Delta g=0.068a$ are plotted in (b) and (c), respectively, with $w_1=0.256a$ and $w_2=0.296a$ in both panels. The directionality $\eta$ of the band in panel (b) and (c) with blue and red color are plotted as blue and red curves in panel (d), respectively. The field pattern of the UGRs marked by the green and purple points in panel (b) and (c) are plotted in panel (e) and (f), respectively. }
\label{figure_1}
\end{figure*}

\subsection{A specific structure with UGRs}

As $w_1=w_2$, because of the center symmetry, a pair of UGRs with the same resonant frequency, opposite $k_x$, and opposite $\eta$ coexist in one structure. As $w_1$ becoming different from $w_2$, the structure breaks the center symmetry, so that the pair of UGRs exist in different structures.
A specific example with $w_1=w_2=0.296a$ and $\Delta g=0.062a$ is given in Fig. 2 of the Ref [27]. As $w_1$ being changed to be $w_1=0.256a$, and $w_2$ remain being $w_2=0.296a$, two structures with $\Delta g=0.053a$ and $\Delta g=0.068a$ have one UGR at $(ak_x)/(2\pi)=0.166$ and $(ak_x)/(2\pi)=-0.111$, respectively, as shown in Fig. 1 (b-d).
In the following, we designate the UGRs with $k_x>0$ or $k_x<0$ as UGR1 and UGR2, respectively. The pair of the two UGRs appear at two different structures. In the other words, for a given structure, only one UGR appear at the part of the resonant band with $k_x>0$ or $k_x<0$.
The field pattern of the UGRs in Fig. 1 (b) and (c) are plotted in Fig. 1 (e) and (f), which only have radiation loss to up-right and down-left direction, respectively. For the structure in Fig. 1 (b), as $\eta$ of UGR1 being unity, $\eta$ of another leaky resonant mode with opposite $k_x$ (the mode at $(ak_x)/(2\pi)=-0.166$) is nearly equal to $-1$. Thus, the leaky resonant mode is a quasi-UGR, which radiate most of the energy to the down side of the meta-grating. For the structure in Fig. 1 (c), the band structure has similar property.

\subsection{Distribution of UGRs in Parameter Space}
When $w_2$ is fixed to be $w_2=0.296a$, and $w_1$ is scanned, the pair of UGRs move in the synthetic parameter space of $(k_x, \Delta g)$ in different directions. For the branches of UGRs corresponding to the scanning of $w_1$, the value of $\Delta g$ and $k_x$ versus $w_1$ are plotted in Fig. 2 (a) and (b), respectively.
The branch of UGR1 only exist within the regime $w_1 \in [0.245a, 0.497a]$, and the branch of UGR2 only exist within the regime $w_1 \in [0.174a, 0.346a]$. As a result, the pair of UGR1 and UGR2 coexist in the synthetic parameter space only when $w_1$ is between $[0.245a, 0.346a]$.
The resonant frequencies and $Q$ factors corresponding to the branches of UGRs are plotted in Fig. 2 (c) and (d), respectively. The resonant frequencies of the UGRs weakly depend on $w_1$. When $w_1$ approaches the lower bound of the branches of UGR1, the $Q$ factor sharply increase to $10^3$, but does not approach infinity. As $w_1$ further decrease, the $Q$ factor of the leaky resonant mode at $\Gamma$ point can further increase as long as the synthetic parameters $(k_x, \Delta g)$ change along the correct direction, but the leaky resonant mode is no longer UGR. When $w_1$ approaches the upper bound of the branches of UGR1, the $Q$ factor does not sharply increase. Similar phenomenon appears for the branches of UGR1.

\begin{figure}[tbp]
\scalebox{0.66}{\includegraphics{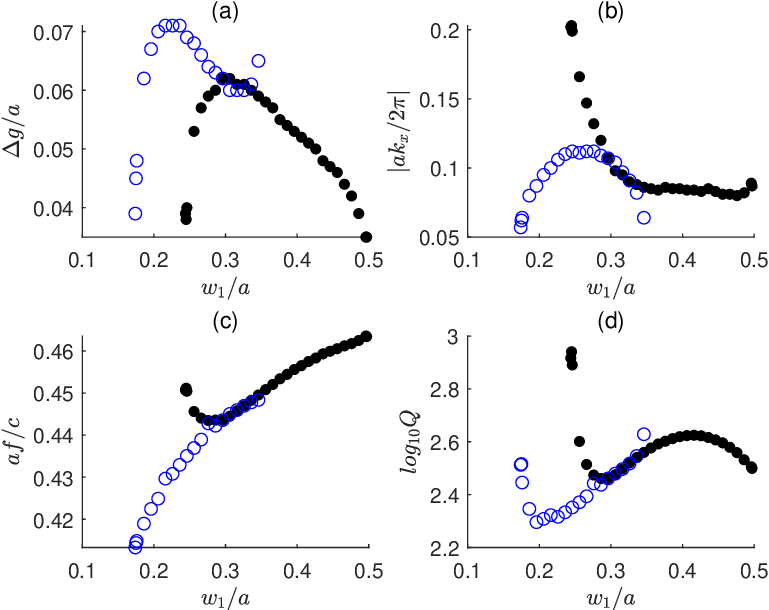}}
\caption{ When $w_2=0.296a$ is fixed, and $w_1$ is scanned, $\Delta g$ of the systems that host UGR with positive and negative $k_x$ versus $w_1$ are plotted as solid black and empty blue dots in panel (a), respectively. The $k_x$, resonant frequency, and $Q$ factor of the corresponding UGR versus $w_1$ are plotted in panel (b), (c) and (d), respectively. }
\label{figure_2}
\end{figure}

\section{GH shifts driven by UGRs}

\subsection{GH Shift of quasi-plane wave incidence}

In this sub-section, we consider an incident quasi-plane wave, which is a Gaussian beam with infinite beam width. The frequency of the incident wave is the same as the resonant frequency of a UGR. The lateral wavenumber of the incident wave is scanned across of the wavenumber of the UGR by changing the incident angle. The angular spectrum of the reflected (transmitted) coefficient, designated as $r$ ($t$), can be obtained by the finite element method simulation of plane wave incident with varying incident angle vie commercially available software COMSOL Multiphysics. The reflectance and transmittance are given as $|r|^2$ and $|t|^2$, respectively, assuming that the amplitude of the incident plane wave is unity; the phase of the reflected and transmitted plane wave is given as $\phi_r = \text{Arg}(r)$ and $\phi_t = \text{Arg}(t)$, respectively.
According to the stationary-phase theory, the GH shifts of the reflective and transmitted fields are given by the change rate of the reflected and transmitted phases versus incident angle, respectively. In another words, the GH shifts are given as  \cite{stationaryPt48}
\begin{equation}
S_{\text{GH},r(t)} = -\frac{\lambda}{2\pi} \frac{\partial \phi_r (\phi_t)}{\partial \theta_{\text{inc}}} \label{GHstationary}
\end{equation}
where $\lambda$ is the wavelength of the incident field, $\theta_{\text{in}}$ is the incident angle.
In previous section, two branches of UGRs are given, whose structural parameters are given in Fig. 2 (a), and the corresponding resonant frequency $f$ and wavenumber $k_x$ are given in Fig. 2 (c) and (b), respectively. For one UGR, as the incident angle being $\theta_r = \arcsin\left( \frac{ck_x}{2\pi f n_b} \right)$, the incident plane wave is resonant with the UGR. As $\theta_{\text{in}}$ being scanned across $\theta_r$, a resonant peak should appear to drive the enhancement of the GH shift. However, the numerical results exhibit anomalous features.

We use the structure in Fig. 1 (b) to demonstrate the feature of the angular spectrum of reflectance (transmittance) and GH shift as the incident wave being resonant with the UGR. Four types of incident conditions are calculated, which are incident from the right-upper, right-lower, left-upper and left-lower background of the meta-grating. The sign convention of the incident angle assumes that incidence from right-upper and right-lower (left-upper and left-lower) background have negative (positive) incident angle, i.e., $\theta_{\text{inc}} < 0$ ($\theta_{\text{inc}} > 0$).
The numerical result of the angular spectrum of reflectance (transmittance), reflected (transmitted) phase and GH shifts are plotted in Fig. 3. The resonant frequency of the UGR1 in Fig. 1 (b) is $af/c=0.4456$, and the corresponding resonant incident angle is $\theta_r=14.95^o$. Thus, in the simulation, the frequency of the incident plane wave is $af/c=0.4456$, and the magnitude of the incident angle is scanned between $[14^o, 16^o]$.
The transmittance of the four types of incident conditions is nearly unity with weak dependent on $\theta_{\text{in}}$, as shown in Fig. 3 (a-d). In another words, the reflectance and transmittance do not have strong resonant peak. However, the phase of the transmitted field has different behavior. When the incident field is from the right-upper and left-lower background, the phases of the transmitted field rapidly change near to the resonant angle of the UGR1, as shown in Fig. 3 (e) and (h), respectively. Thus, the GH shift of the transmitted field has a resonant peak, as shown in Fig. 3 (i) and (l). Because of time reversal symmetry, the GH shift of the transmitted field of the two types of incidences are the same.
On the other hand, when the incident field is from the right-lower and left-upper background, the phases of the transmitted field are nearly constant, as shown in Fig. 3 (f) and (g), respectively. Thus, the GH shifts of the corresponding transmitted field are nearly zero, as shown in Fig. 3 (j) and (k). 

\begin{figure*}[tbp]
\scalebox{0.53}{\includegraphics{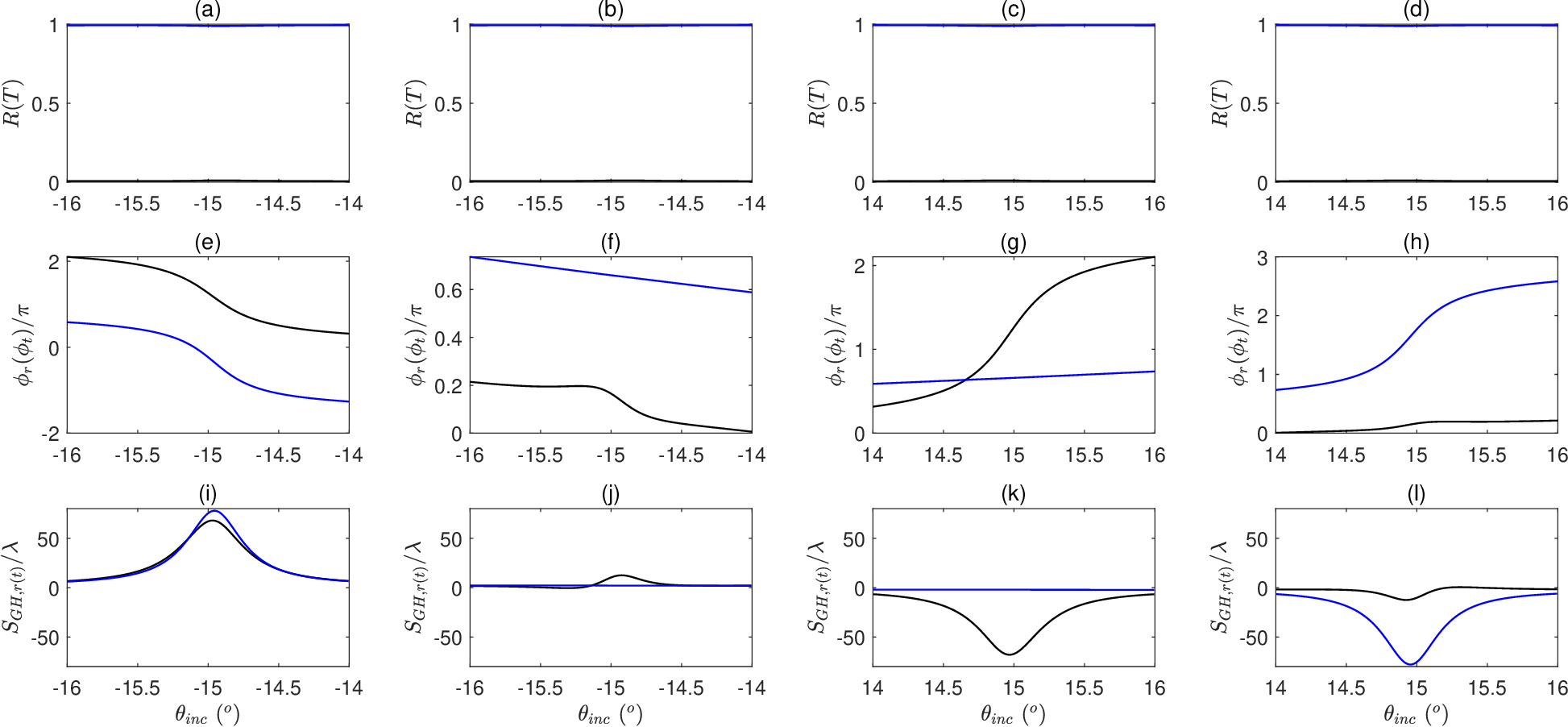}}
\caption{ For the structure in Fig. 1 (b), the angular spectrum of reflectance and transmittance are plotted as black and blue lines in the first row, respectively; the angular spectrum of the phase of the reflected and transmitted field are plotted as black and blue lines in the second row, respectively; the GH shift given by the stationary-phase theory of the reflected and transmitted field are plotted as black and blue lines in the third row, respectively. For the first to fourth column, the incident plane waves are incident from the right-upper, right-lower, left-upper and left-lower background of the meta-grating.  }
\label{figure_3}
\end{figure*}

If the structural parameters, incident frequency and angle of the systems are chosen to be corresponding to the other UGRs with parameters being given in Fig. 2, the features of the angular spectrum are the same, but the maximum magnitude of the GH shift of the transmitted field with incident field from right-upper and left-lower background are different. The maximal GH shift of the UGRs versus the corresponding structural parameter are plotted in Fig. 4 (a).
For some of the UGRs, the GH shift is linearly proportional to the $Q$ factor; for the others UGRs, the GH shifts are suppressed, as shown in Fig. 4 (b). For the branches of UGR1, as $w_1$ approaching the lower bound at $0.245a$, the GH shift sharply increases, which is proportional to the sharply increase of the $Q$ factor; as $w_1$ being larger than $0.3a$, the GH shift is suppressed to be smaller than $40\lambda$.
For the branches of UGR2, as $w_1/a \in [0.236, 0.296]$, the GH shift is proportional to the $Q$ factor; as $w_1$ approaching the lower bound at $0.174a$, the GH shift is suppressed to be near zero, although the $Q$ factor is larger than 300. The anomalous suppression of the GH shift can be explained by the TCMT analysis in the next subsection.

\begin{figure}[tbp]
\scalebox{0.66}{\includegraphics{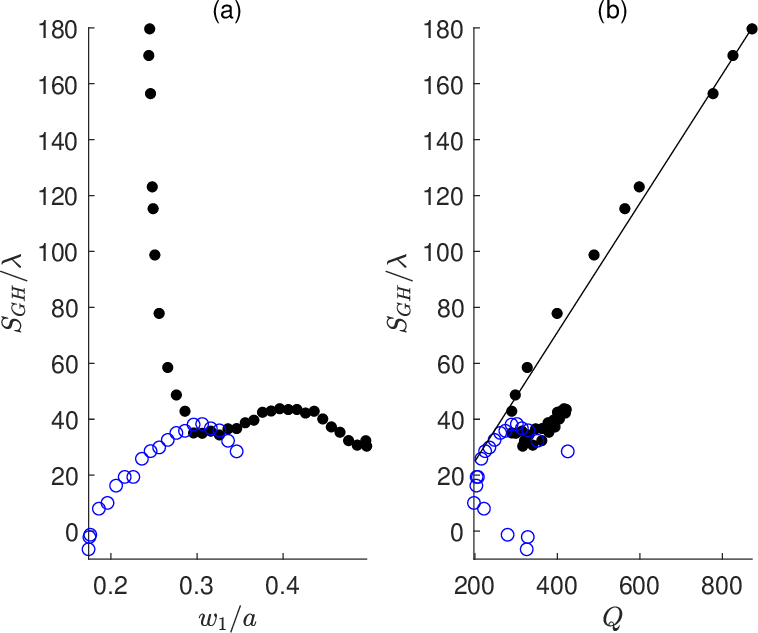}}
\caption{ (a) The maximal GH shift in the angular spectral versus the structural parameter $w_1$ of the branches of UGRs with positive and negative $k_x$ are plotted as solid black and empty blue dots, respectively. (b) The same as panel (a) with the x axis being replaced by the corresponding $Q$ factors of the UGRs. The solid thin line fits the proportional relation between the GH shift and the $Q$ factor.  }
\label{figure_4}
\end{figure}

\subsection{TCMT analysis}

For a given magnitude of the transversal wavenumber $|k_x|$, the resonant frequency of a leaky resonant mode at the band structure is designated as $\omega_b$, and the decay rate of the leaky resonant mode is designated as $\gamma_b$. The value of $\omega_b$ and $\gamma_b$ and given by the real and imaginary part of the eigen-frequency. Because of time-reversal symmetry, the eigen-frequencies of the two resonant modes with opposite sign of $k_x$ are the same. For the UGR in the band structure, the corresponding $|k_x|$, $\omega_b$ and $\gamma_b$ are designated as $k_U$, $\omega_U$ and $\gamma_U$, respectively.
When an incident plane wave with frequency being $\omega$ and incident angle being $\theta_{\text{inc}} = \pm \arcsin\left(\frac{|k_x|c}{\omega}\right)$ from the upper or lower backgrounds, the two leaky resonant modes with wavenumber being $\pm|k_x|$ and resonant frequency being $\omega_b$ can be excited. The scattering channels to the right-upper, right-lower, left-upper and left-lower directions of the meta-grating are designated as channels 1, 2, 3 and 4, respectively. The incident field from the four channels are designated as $|s^{(+)}\rangle = [s_1^{(+)}, s_2^{(+)}, s_3^{(+)}, s_4^{(+)}]^T$, and the outgoing field to the four channels are designated as $|s^{(-)}\rangle = [s_1^{(-)}, s_2^{(-)}, s_3^{(-)}, s_4^{(-)}]^T$. The time-dependent amplitude of the leaky resonant modes with wavenumber being $-|k_x|$ and $|k_x|$ are designated as $a_1(t)$ and $a_2(t)$, which are written as $|a(t)\rangle = [a_1(t), a_2(t)]^T$.
Because of the preservation of the transversal momentum, the incident field from scattering channels one and two can only excite the mode with amplitude being $a_1(t)$, and generate scattering field to channel three and four; similarly, the incident field from scattering channels three and four can only excite the mode with amplitude being $a_2(t)$, and generate scattering field to channel one and two. Thus, the coupling coefficient between the incident fields and the two resonant modes are designated as
\begin{equation}
|\kappa\rangle = \begin{bmatrix}
\begin{matrix} \kappa_1 & 0 \\ \kappa_2 & 0 \end{matrix} \\
\begin{matrix} 0 & \kappa_3 \\ 0 & \kappa_4 \end{matrix}
\end{bmatrix} \label{tcmt02}
\end{equation}
so that the rate of excitation of the resonant modes by the incident fields is given as $\langle\kappa^*|s^{(+)}\rangle$, with $\langle\kappa^*|$ being the transpose of $|\kappa\rangle$. According to the TCMT, the amplitude of the resonant modes are given as \cite{tcmt01}
\begin{equation}
\frac{d}{dt}|a(t)\rangle = (-i\omega_b - \gamma_b)|a(t)\rangle + \langle\kappa^*|s^{(+)}\rangle \label{tcmt03}
\end{equation}
If the incident field is monochromatic plane wave with frequency being $\omega$, the amplitude of the resonant modes can be written as $|a(t)\rangle = |a\rangle e^{-i\omega t}$, and then the solution of Eq. (\ref{tcmt03}) is given as
\begin{equation}
|a\rangle = \frac{\langle\kappa^*|s^{(+)}\rangle}{-i(\omega - \omega_b) + \gamma_b} \label{tcmt04}
\end{equation}
Because the mode with amplitude being $a_1$ ($a_2$) can only radiate energy to channel three and four (one and two), the radiation from the resonant modes is written as $|d\rangle|a\rangle$, with $|d\rangle$ being the coupling coefficient between the outgoing fields and the two resonant modes, which is given as
\begin{equation}
|d\rangle = \begin{bmatrix}
\begin{matrix} 0 & d_1 \\ 0 & d_2 \end{matrix} \\
\begin{matrix} d_3 & 0 \\ d_4 & 0 \end{matrix}
\end{bmatrix} \label{tcmt05}
\end{equation}
Because of time-reversal symmetry, we have $d_i = \kappa_i$ with $i \in [1,2,3,4]$. Because of energy conservation, the coefficient $|d\rangle$ satisfies the relation $\langle d|d\rangle = 2\gamma_b I$ with $I$ being a two-by-two unit matrix, i.e., $2\gamma_b = |d_1|^2 + |d_2|^2 = |d_3|^2 + |d_4|^2$. The value of $d_1/d_2$ ($d_3/d_4$) is equal to the value of $c_{\text{up}}/c_{\text{down}}$ of the corresponding resonant modes.
In addition to radiation from the resonant modes, the incident fields also experience direct scattering, which is described by the scattering matrix
\begin{equation}
\hat{C} = \begin{bmatrix}
\begin{matrix} 0 & 0 \\ 0 & 0 \end{matrix} & \begin{matrix} C_{11} & C_{12} \\ C_{21} & C_{22} \end{matrix} \\
\begin{matrix} C_{11} & C_{21} \\ C_{12} & C_{22} \end{matrix} & \begin{matrix} 0 & 0 \\ 0 & 0 \end{matrix}
\end{bmatrix} \label{tcmt06}
\end{equation}
Because of time reversal symmetry, $\hat{C}_{13} = \hat{C}_{31}$, $\hat{C}_{24} = \hat{C}_{42}$, $\hat{C}_{14} = \hat{C}_{41}$, and $\hat{C}_{23} = \hat{C}_{32}$, so that the corresponding matrix elements have the same notation. Because of energy conservation, for dielectric meta-grating without material absorption loss, $|C_{11}|^2 + |C_{21}|^2 = |C_{11}|^2 + |C_{12}|^2 = |C_{22}|^2 + |C_{21}|^2 = |C_{22}|^2 + |C_{12}|^2 = 1$. The scattered field given by the direct scattering process is $\hat{C}|s^{(+)}\rangle$. As a result, the total scattered field is given as
\begin{equation}
|s^{(-)}\rangle = \left( \hat{C} + \frac{|d\rangle\langle\kappa^*|}{-i(\omega - \omega_b) + \gamma_b} \right) |s^{(+)}\rangle \label{tcmt07}
\end{equation}
If $\omega$ is fixed, the angular spectrum or reflectance (transmittance) and reflected (transmitted) phase versus $\theta_{\text{inc}}$ can be modeled by Eq. (\ref{tcmt07}). Because of energy conservation, we have
\begin{equation}
\hat{C}|d^*\rangle = -|\kappa\rangle \label{tcmt08}
\end{equation}
with $|d^*\rangle$ being complex conjugate of $|d\rangle$. The value of each term in $\hat{C}$ can be obtained by solving Eq. (\ref{tcmt08}).

We firstly consider the structures with $w_1=w_2$, which have both UGR1 or UGR2 with opposite $k_x$. The amplitude of the UGR1 or UGR2 are $a_2$ and $a_1$, respectively. The frequency of the incident field is chosen to be resonant with the UGRs, i.e., $\omega=\omega_U$. If the incident angle is $\theta_{\text{inc},U}=\arcsin\left(\frac{k_U c}{\omega n_b}\right)$ ($\theta_{\text{inc},U}=-\arcsin\left(\frac{k_U c}{\omega n_b}\right)$), and the incident field is from the lower (upper) background, the UGR1 (UGR2) is excited.
According to the numerical result in Fig. 1, the UGR1 and UGR2 only have radiative loss to right-upper and left-lower background (channel one and four), so that $d_2=d_3=0$, and $d_1=d_4=\sqrt{2\gamma_b}$. By applying Eq. (\ref{tcmt08}), we have $C_{11}=C_{22}=0$, $C_{12}=-1$ and $C_{21}=e^{i\varsigma}$, with $\varsigma$ being an arbitrary phase factor.
If the incident angle deviates from $\pm \arcsin\left(\frac{k_U c}{\omega}\right)$, the leaky resonant modes in the band structure near to the UGRs are excited, so that the value of $\omega_b$, $\gamma_b$ and $|d\rangle$ change. Numerical results show that the value of each term of $|d\rangle$ smoothly changes versus $k_x$, so that in the narrow range of incident angle, $|d\rangle$ can be considered as constant. Thus, the value of the matrix elements of $\hat{C}$ is also constant.
The eigen-frequency of the band structure near to the UGR can be given by first order Taylor expansion, i.e., $\omega_b=\omega_U+\delta \omega_{U,1} (k_x-\delta k_U )$ and $\gamma_b=\gamma_U+\delta \gamma_{U,1} (k_x-\delta k_U )$, with $\omega_{U,1}=\left. \frac{\partial \omega_b}{\partial k_x} \right|_{k_x=k_U }$, $\gamma_{U,1}=\left. \frac{\partial \gamma_b}{\partial k_x} \right|_{k_x=k_U }$, and $\delta=\pm1$ for the cases with $k_x>0$ ($k_x<0$).
Inserting the value of $|d\rangle$, $\hat{C}$ and $\omega_b$ into Eq. (\ref{tcmt07}), the scattering field to the four channels are given as
\begin{eqnarray}
&&s_1^{(-)} = \label{tcmt09}\\
&&\left[ -1 + \frac{2\left( \gamma_U + \delta \gamma_{U,1} (k_x - \delta k_U ) \right)}{i\delta \omega_{U,1} (k_x - \delta k_U ) + \gamma_U + \delta \gamma_{U,1} (k_x - \delta k_U )} \right] s_4^{(+)} \nonumber
\end{eqnarray}
\begin{equation}
s_2^{(-)} = e^{i\varsigma} s_3^{(+)} \label{tcmt10}
\end{equation}
\begin{equation}
s_3^{(-)} = e^{i\varsigma} s_2^{(+)} \label{tcmt11}
\end{equation}
\begin{eqnarray}
&&s_4^{(-)} = \label{tcmt12}\\
&&\left[ -1 + \frac{2\left( \gamma_U + \delta \gamma_{U,1} (k_x - \delta k_U ) \right)}{i\delta \omega_{U,1} (k_x - \delta k_U ) + \gamma_U + \delta \gamma_{U,1} (k_x - \delta k_U )} \right] s_1^{(+)} \nonumber
\end{eqnarray}
Because of conservation of $k_x$ in the scattering process, $\delta=+1$ in Eq. (\ref{tcmt09})-(\ref{tcmt10}), and $\delta=-1$ in Eq. (\ref{tcmt11})-(\ref{tcmt12}). For the incidences from the four channels, the reflectance is zero. For the incidences from channels two and three, the transmission coefficients are constant, as shown in Eq. (\ref{tcmt11}) and (\ref{tcmt10}), respectively, so that the GH shift is zero. For the incidences from channels one and four, the transmission coefficients in Eq. (\ref{tcmt12}) and (\ref{tcmt09}) can be simplified to be $t_\delta = \frac{-i\delta \omega_{U,1} (k_x - \delta k_U ) + \gamma_U + \delta \gamma_{U,1} (k_x - \delta k_U )}{i\delta \omega_{U,1} (k_x - \delta k_U ) + \gamma_U + \delta \gamma_{U,1} (k_x - \delta k_U )}$ with $\delta=-1$ and $\delta=1$, respectively.
The absolute value of $t_\delta$ is unity without resonant peak, but the complex phase of $t_\delta$ has a resonant peak. With $k_x = \frac{\omega_U n_b}{c} \sin\theta_{\text{inc}}$, the angular spectrum of the complex phase of $t_\delta$ is given as
\begin{equation}
\phi = 2\arctan\left( \frac{-\delta \omega_{U,1} \left( \frac{\omega_U n_b}{c} \sin\theta_{\text{inc}} - \delta k_U \right)}{\gamma_U + \delta \gamma_{U,1} \left( \frac{\omega_U n_b}{c} \sin\theta_{\text{inc}} - \delta k_U \right)} \right) \label{tcmt13}
\end{equation}
Because $\theta_{\text{inc}}$ is scanned across a narrow range near $\delta \theta_{\text{inc},U}$ with $\theta_{\text{inc},U}=\arcsin\left( \frac{k_U c}{\omega_U n_b } \right)$, by approximating the sine function as first order Taylor expansion near $\delta \theta_{\text{inc},U}$, $\sin\theta_{\text{inc}} \approx \frac{\delta k_U c}{\omega_U n_b } + \cos\theta_{\text{inc},U} \cdot (\theta_{\text{inc}} - \delta \theta_{\text{inc},U})$, so that $\phi \approx 2\arctan\left( \frac{-\delta \omega_{U,1} \omega_U n_b \cos\theta_{\text{inc},U} \cdot (\theta_{\text{inc}} - \delta \theta_{\text{inc},U})}{c\gamma_U + \delta \gamma_{U,1} \omega_U n_b \cos\theta_{\text{inc},U} \cdot (\theta_{\text{inc}} - \delta \theta_{\text{inc},U})} \right)$. The GH shift is given as
\begin{eqnarray}
&&S_{\text{GH},t_\delta} = -\frac{\lambda}{2\pi} \frac{\partial \phi}{\partial \theta_{\text{inc}}} \label{tcmt14}\\
&&= \frac{\lambda}{\pi} \cdot \frac{\delta c \gamma_b \omega_{U,1} \omega_U n_b \cos\theta_{\text{inc},U}}{c^2 \gamma_b^2 + \left( \omega_{U,1} \omega_U n_b \cos\theta_{\text{inc},U} \right)^2 (\theta_{\text{inc}} - \delta \theta_{\text{inc},U})^2} \nonumber
\end{eqnarray}
As $\theta_{\text{inc}}=\delta \theta_{\text{inc},U}$, the GH shift is given as
\begin{equation}
S_{\text{GH},t} = \frac{\delta \lambda \omega_{U,1} \omega_U n_b \cos\theta_{\text{inc},U}}{\pi c \gamma_U} \label{tcmt15}
\end{equation}
so that the GH shift is determined by the value of $\omega_{U,1}$ and $\frac{\omega_U}{\gamma_U}$. $\omega_{U,1}$ is equal to the group velocity of the UGR, and $\frac{\omega_U}{\gamma_U}$ is the $Q$ factor of the UGR. Thus, the GH shift is proportional to the group velocity as well as the $Q$ factor of the UGR.
For the UGR1 and UGR2, the group velocity is positive and negative, so that the GH shift of the UGR1 and UGR2 are positive and negative, respectively. Numerical results show that the group velocity is approximately equal to $\frac{c}{n_b}$, so that the GH shift is solely proportional to the $Q$ factor. When $|\theta_{\text{inc}} - \delta \theta_{\text{inc},U}| = \frac{c\gamma_b}{\omega_{U,1} \omega_U n_b \cos\theta_{\text{inc},U}}$, the GH shift is decreased by half, so that the line width of the GH shift is proportional to $\gamma_b$.
Because the first and second term inside the bracket of Eq. (\ref{tcmt09}) and (\ref{tcmt12}) are due to the direct scattering process and the radiative loss from the UGR, respectively, the anomalous behavior that the peak of GH shift coexists with constant transmittance is driven by the interference between the direct scattering process and resonant excitation of the UGR. 

Secondly, the structures with $w_1 \neq w_2$, which host either UGR1 or UGR2, are considered. For the systems with only UGR2, $d_3$ remains being zero, so that $d_4$ remains being $\sqrt{2\gamma_b}$; $|d_1|$ become slightly smaller than $\sqrt{2\gamma_b}$, and $|d_2|$ becomes slightly larger than zero. By applying Eq. (\ref{tcmt08}), we have $C_{11} = \frac{d_2 e^{i\varsigma_1}}{\sqrt{2\gamma_b}}$, $C_{22} = -\frac{d_2}{\sqrt{2\gamma_b}}$, $C_{12} = -\frac{d_1}{\sqrt{2\gamma_b}}$ and $C_{21} = \frac{d_1 e^{i\varsigma_2}}{\sqrt{2\gamma_b}}$, with $\varsigma_1$ and $\varsigma_2$ being arbitrary phase factors. The solution of $|s^{(-)}\rangle$ is given as
\begin{eqnarray}
&&s_1^{(-)} = \frac{d_2 e^{i\varsigma_1}}{\sqrt{2\gamma_b}} s_3^{(+)} +\label{tcmt16}\\ &&\left[ -\frac{1}{\sqrt{2\gamma_b}} + \frac{\sqrt{2\gamma_b}}{i\delta \omega_{U,1} (k_x - \delta k_U) + \gamma_b} \right] d_1 s_4^{(+)} \nonumber
\end{eqnarray}
\begin{eqnarray}
&&s_2^{(-)} = \frac{d_1 e^{i\varsigma_2}}{\sqrt{2\gamma_b}} s_3^{(+)} + \label{tcmt17}\\ &&\left[ -\frac{1}{\sqrt{2\gamma_b}} + \frac{\sqrt{2\gamma_b}}{i\delta \omega_{U,1} (k_x - \delta k_U) + \gamma_b} \right] d_2 s_4^{(+)} \nonumber
\end{eqnarray}
\begin{equation}
s_3^{(-)} = \frac{d_2 e^{i\varsigma_1}}{\sqrt{2\gamma_b}} s_1^{(+)} + \frac{d_1 e^{i\varsigma_2}}{\sqrt{2\gamma_b}} s_2^{(+)} \label{tcmt18}
\end{equation}
\begin{eqnarray}
s_4^{(-)} = &&\left[ -\frac{1}{\sqrt{2\gamma_b}} + \frac{\sqrt{2\gamma_b}}{i\delta \omega_{U,1} (k_x - \delta k_U) + \gamma_b} \right] \label{tcmt19}\\ &&\left( d_1 s_1^{(+)} + d_2 s_2^{(+)} \right) \nonumber
\end{eqnarray}
According to Eq. (\ref{tcmt16})-(\ref{tcmt17}), for the incident field from channel four, the reflectance and transmittance are proportional to $|d_2|^2$ and $|d_1|^2$, respectively, which are not dependent on $k_x$ or $\theta_{\text{inc}}$. However, the complex phases of the reflected and transmitted coefficient have large slope near $\delta \theta_{\text{inc},U}$. Thus, the GH shifts are nonzero, which are also given by Eq. (\ref{tcmt14}). For the incident field from channel three, the reflected and transmitted coefficient are constant, so that the GH shift is zero.
Similarly, according to Eq. (\ref{tcmt18})-(\ref{tcmt19}), for the incident field from channel one or two, the scattering field to the channel three has no GH shift, and the scattering field to the channel four has GH shift given by Eq. (\ref{tcmt14}). For the structures with parameters given in Fig. 2, the value of $|d_2|$ is much smaller than $|d_1|$, so that the transmittance is near unity. For most of the structures, the GH shift is proportional to the $Q$ factor of the UGR2. However, when $w_1$ approaches the lower bound at $0.174a$, the UGR2 moves to the dip of the band structure, so that the group velocity decreases to zero. As a result, according to Eq. (\ref{tcmt14}), the GH shift is anomalously suppressed to zero. For the systems with only UGR1, similar formula can be obtained.

\subsection{GH shift of Gaussian beam incidence}
In order to exhibit the GH shift in more intuitive way, the GH shift of the meta-grating under the incident of Gaussian beam with finite beam width is calculated. The results are shown in Fig. 5. As the incident field is from the right-upper and left-lower background, the GH shift is large. The field pattern shows that the incident field excites the localized field of the UGR. After travelling the distance that is equal to the GH shift, the localized field radiate the energy to the opposite side of the incident direction, as shown in Fig. 5 (a) and (d). The GH shift is as large as the beam width of the Gaussian beam, so that the incident Gaussian beam and the transmitted Gaussian beam are spatially separated. On the other hand, as the incident field being from the right-lower and left-upper background, the UGR is not excited. The incident field directly transmits to the opposite side of the meta-grating without any GH shift, as shown in Fig. 5 (b) and (c), respectively.

\begin{figure}[tbp]
\scalebox{0.53}{\includegraphics{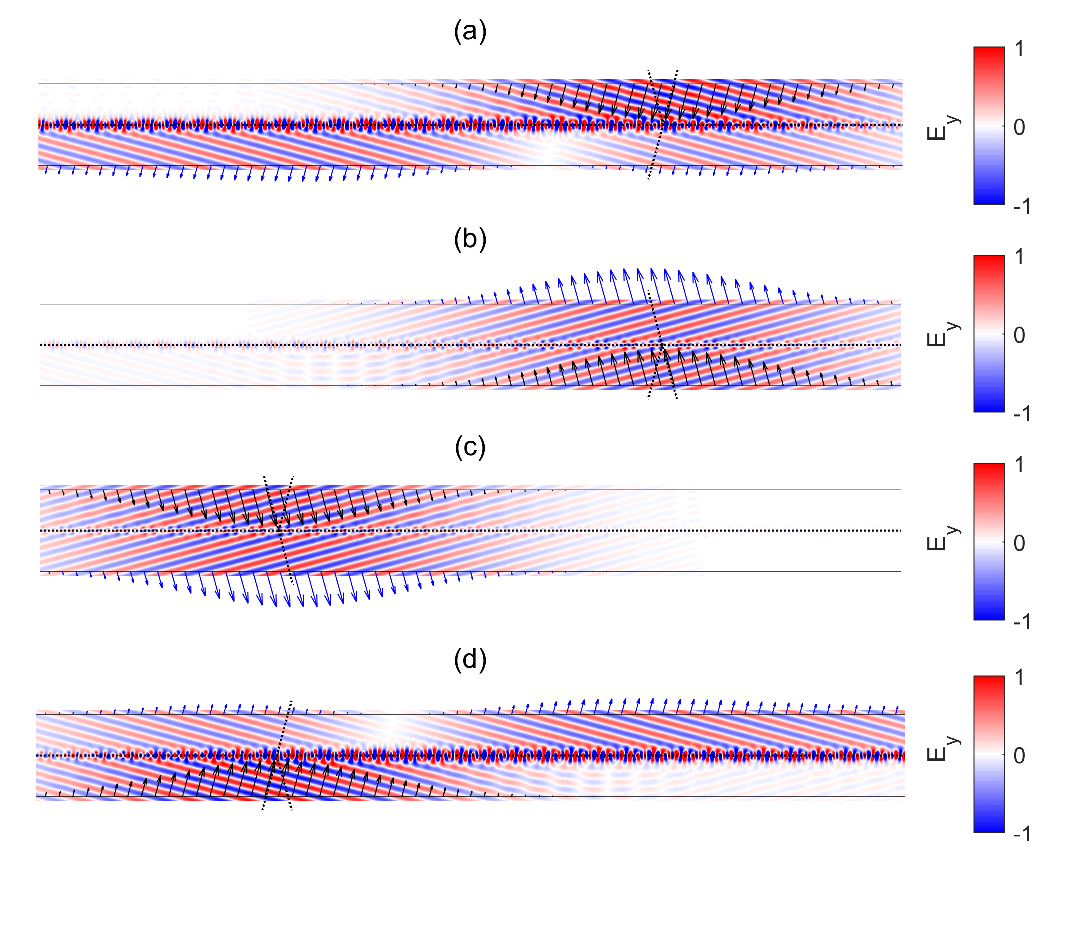}}
\caption{ Spatial distributions of time-averaged Poynting vectors for incident and transmitted fields under the incident Gaussian beams with incident angles being $\theta_{\text{inc}}=\pm14.9502^o$, are plotted as black and blue arrows, respectively. The field pattern of $E_y$ is plotted as color-scale. The frequency is $af/c=0.445614$, and the structural parameters are the same as the system in Fig. 1 (b). The incident field is from the right-upper, right-lower, left-upper and left-lower background of the meta-grating for the cases in (a), (b) (c) and (d), respectively.  }
\label{figure_5}
\end{figure}

\section{conclusion}

The distribution of UGRs in the parameter space of the asymmetric double-layer meta-grating without center inversion symmetry are numerically calculated. The GH shifts of the structures that host UGRs exhibit anomalous behaviors. Firstly, the resonant peak of the GH shift coexists with a constant transmittance, which is due to the interference between the direct scattering process and resonant excitation of the UGR. Secondly, the GH shift is proportional to the group velocity as well as the Q factor of the UGR, so that the GH shift of the UGR with small group velocity is suppressed. Simulation of the Gaussian beam incidence with finite beam width indicate that the GH shift can be as large as the sum of beam widths of the incident and transmitted beams, so that sizable shift of the beam can be applied for photonic devices.

\clearpage

\end{document}